\documentstyle[aps,prb,multicol,psfig]{revtex}
\newcommand{\beq}{\begin{equation}}
\newcommand{\eeq}{\end{equation}}
\newcommand{\bea}{\begin{eqnarray}}
\newcommand{\eea}{\end{eqnarray}}
\newcommand{\ini}{{\mbox{\small i}}}
\newcommand{\fin}{{\mbox{\small f}}}
\newcommand{\inc}{{\mbox{\small -}}}

\newcommand{\vR}{{\bf R}}
\newcommand{\vK}{{\bf K}}
\newcommand{\vp}{{\bf p}}
\newcommand{\vk}{{\bf k}}
\renewcommand{\vr}{{\bf r}}
\renewcommand{\narrowtext}{\begin{multicols}{2} \global\columnwidth20.5pc}
\renewcommand{\widetext}{\end{multicols} \global\columnwidth42.5pc}

\begin{document}
\draft
\title{Double photo-ionization of He near a polarizable surface}

\author{Guido Goldoni}

\address{Istituto Nazionale per la Fisica della Materia (INFM), and \\
Dipartimento di Fisica, Universit\`a di Modena, Via Campi
213/A, I-41100 Modena, Italy}

\date{\today}
\maketitle

\begin{abstract}

We calculate the differential cross-section of the direct double
photo-ionization of He physisorbed on a polarizable surface.
By including the influence of the surface potential in the correlated
two-electron final state wavefunction, we show that the differential
cross-section carries detailed information on the electronic
correlations at the surface. In particular, photo-emission along
opposite directions, which is prohibited in the free space, is allowed
if the surface potential is long-ranged.

\end{abstract}


\narrowtext

The simultaneous photo-emission of two electrons $e_1$ and $e_2$
(direct double photo-ionization, DPI) is a straight manifestation of
electronic correlations in the initial and/or final state.
Despite this fundamental aspect, the experimental determination of the
differential cross-section (DCS) of this process and quantitative
agreement with theoretical calculations are relatively recent
achievements even in the simplest atomic systems\cite{DCS-He}. 
Stimulated by the present availability of bright synchrotron 
radiation sources, it has now become
of interest to look at the DPI process also in solid-state systems\
\cite{Berakdar98}, in order to investigate electronic correlations in the
material, particularly near the surface.

As a step in this direction, in this paper we study the angular
distribution of photo-electrons in a DPI process 
in the simplest two-electron atomic system, an He
atom, adjacent to a polarizable surface. By virtue of the dipole
interaction with the surface, at low temperature an He atom weakly
adheres to the surface (physisorption) at a rather large 
distance $d$ from the surface edge.
The He (correlated) ground state is only sligthly distorted by the
weak atom-surface iteraction; therefore, possible effects due to the
surface should primarily be regarded as final state effects.
In fact, we shall demonstrate that the DCS of the DPI process from
physisorbed He is affected by the nearby surface and, in
particular, atomic selection rules are relaxed in a way that
depends on the surface potential. Therefore, the measurement of the
DPI cross-section from physisorbed He might give detailed
informations on the surface potential, in analogy with the angle-resolved
spectroscopy of image states\cite{Osgood97}.

In the velocity gauge \cite{gauge}, the DCS of the DPI process is 
\beq
\frac{d\sigma}{d\Omega_1 d\Omega_2 dE} \propto
\left|\langle\Psi_\ini|\bbox{\hat{\varepsilon}}\cdot(\vp_1+\vp_2)
|\Psi^{\inc}_\fin\rangle\right|^2,
\label{dpi_cs}
\eeq
where $\bbox{\hat{\varepsilon}}$ is the light polarization vector,
$\vp_i$ is the momentum operator for $e_i$, and
$\Psi_{\ini}$ and $\Psi^{\inc}_\fin$ are the two-electron initial and
final state, respectively.  $\Omega_i$ denotes the solid angle at which
$e_i$ is detected with energy $E_i$, and
$E=E_1+E_2$.

We next consider the case of equal energy
sharing, $E_1=E_2=E/2$, between the photo-electrons. In this case,
the selection rules\cite{DCS-He}
\begin{mathletters}
\bea
\vk_1+\vk_2 & \neq & 0 \label{sel_rule_1}, \\
\vk_1-\vk_2 & \neq & 0 \label{sel_rule_2},
\eea
\end{mathletters}
apply for free He, i.e., 
photo-electrons cannot be emitted either in the same or in opposite
directions and with the same energy \cite{UES}. 
Equation (\ref{sel_rule_1}) derives from the form of the
operator $\vp_1+\vp_2$: the DCS
is zero if the final state corresponds to a channel with the
photo-electrons emitted with zero center-of-mass momentum.
This is
easily seen if $|\Psi^{\inc}_\fin\rangle$ is taken as the
uncorrelated product of two plane waves with wavevectors $\vk_1,
\vk_2$, but Eq.\ (\ref{sel_rule_1}) holds also when more realistic
final states are used, with the $e_1$-$e_2$ repulsion correctly
taken into account; it is such repulsion which leads to the
second selection rule, Eq.\ (\ref{sel_rule_2}). 
Eqs.\ (\ref{sel_rule_1}),(\ref{sel_rule_2}) have been 
experimentally confirmed for free He\cite{DCS-He}.

Obviously, a surface breaks the translational invariance in one
direction and we
may argue that the DCS has a different shape in this case. The system we are mainly
interested in is the surface of a metal, treated here within the
jellium model. We shall demonstrate by explicit calculations that for
physisorbed He i) the selection rule (\ref{sel_rule_1}) is relaxed
when the long-ranged potential outside the jellium surface is taken
into account; ii) if, on the other hand, the surface potential is
characterized by a screening length $l_0$, atomic selection rules are
recovered if $l_0$ is comparable to or smaller than the physisorbion
distance $d$; in this
regime, on the other hand, the DCS becomes peaked along particular directions.

In proximity of a polarizable surface, the Coulomb interaction between
two charges is renormalized with respect to vacuum and, in general, it
is also dependent on the dynamical state ---energy and momentum--- of
the interacting particles. This renormalization affects both the
$e_1$-$e_2$ interaction, $V_{12}$, and the electron-ion core
interaction, $V_{\mbox{\tiny ion},i}$; in addition, the
renormalization of the Coulomb interaction gives rise to a
single-particle potential (surface potential) $\Sigma_i(z_i)$ for
$e_i$; outside a metallic surface, $\Sigma_i(z_i)$ is long-ranged, and
coincides with the classical image potential $-1/4z_i$ at large $z_i$
(we consider a surface filling the half-space $z<0$; energy is
measured with respect to vacuum; hartree atomic units are used
throughout), while it converges to the inner potential inside the
metal (see inset of Fig.\ 1)\cite{Eriksson85,deAbajo92}. The final
state two-electron Hamiltonian $H_\fin$ can be written as
\bea
H_\fin & = & H_1 + H_2 + V_{12}, \\
H_i & = & -\frac{1}{2}p_i^2 + \Sigma_i + V_{\mbox{\tiny ion},i}, 
\label{H}
\eea
where the single-particle Hamiltonian $H_i$ acts on the coordinates 
$\vr_i$ of $e_i$. Because of the large ionic mass, $V_{\mbox{\tiny ion},i}$ is
simply the potential of a dipole comprising the ion core and its
classical image charge and, therefore, it is short-ranged. Accordingly,
we neglect $V_{\mbox{\tiny ion},i}$ with respect to $\Sigma_i$, and we
define an approximate single-particle Hamiltonian $ \tilde{H}_i
=-\frac{1}{2}p_i^2 + \Sigma_i $ as the limit of $H_i$ at large
distances from the surface. 

To calculate the two-electron eigenstates of the approximate final state
Hamiltonian $\tilde{H}_\fin  = \tilde{H}_1 + \tilde{H}_2 + V_{12} $ we
consider wavefunctions of the form 
\beq
\langle\vr_1,\vr_2|\Psi^{\inc}_\fin\rangle = 
\psi^{\inc}_{\alpha_1} (\vr_1) \psi^{\inc}_{\alpha_2} (\vr_2) 
M^{\inc}_{\alpha_1 \alpha_2}(\vr).
\label{psi_fin}
\eeq
where 
$\psi^{\inc}_{\alpha_i}(\vr_i)$ is a scattering state of $ \tilde{H}_i $
with quantum numbers $\alpha_i$ and incoming boundary conditions;
$M^{\inc}_{\alpha_1 \alpha_2}(\vr)$ 
is assumed to depend only on the relative coordinates
$\vr=\vr_1-\vr_2$ and is the solution of a differential equation, to
be discussed below, ensuing from the
Schr\"odinger equation with the Hamiltonian $\tilde{H}_\fin$\cite{nota3C}
and the {\em ansatz} (\ref{psi_fin}).  

In the following we distinguish between in-plane and $z$ components of
the coordinates, with the definitions $\vr_i = (\vR_i,z_i)$ and $\vr =
(\vR,z)$. Due to the translational invariance parallel to the surface
we can write $\psi^{\inc}_{E_i,\vK_i}(\vr_i) = e^{i
\vK_i\cdot\vR_i}\phi^*_{E_i,\vK_i}(z_i)$, where $\vK_i$ is the in-plane
wavevector of $e_i$, and $\phi_{E_i,\vK_i}(z_i)$ is the
solution of the equation
\beq
\left[-\frac{1}{2}\frac{\partial^2}{\partial z_i^2}+
\Sigma(z_i)+\frac{K_i^2}{2}-E_i\right]\phi_{E_i,\vK_i}(z_i)=0,
\label{sp}
\eeq
which can be solved by direct integration, with the boundary condition
that deep inside the metal $\phi_{E_i,\vK_i}(z_i)$
becomes a plane-wave propagating toward the left.

\widetext
The function $M^{\inc}_{E_1 \vK_1 E_2 \vK_2}(\vr)$ is the solution of 
\beq
\left\{-\nabla_\vR^2 - \frac{\partial^2}{\partial z^2} 
-i\left[\vK\cdot\nabla_\vR + F(z_1,z_2) 
\frac{\partial}{\partial z} \right] + 
V_{12}(\vR,z)\right\} M^{\inc}_{E_1 \vK_1 E_2 \vK_2}(\vR,z) = 0, 
\label{diffeq}
\eeq
where $\vK=\vK_1-\vK_2$, and we have defined 
$F(z_1,z_2)=f(\phi_{\alpha_1},z_1) - f(\phi_{\alpha_2},z_2)$, with  
$f(\phi_{\alpha_i},z_i)=
\phi_{\alpha_i}^{-1} \partial \phi_{\alpha_i} /\partial z_i $.
Note that, due to the presence of the surface potential,
$M^{\inc}(\vR,z)$ depends parametrically on the center of mass of the
two electrons through the function $F(z_1,z_2)$; in the
following we take $F(z_1,z_2) = -i(k_{z1}-k_{z2})$, 
with $k_{zi} = \sqrt{2(E_i-K_i^2/2)}$; this is the value
of $F(z_1,z_2)$ if $\phi_{E_i,\vK_i}(z_i)$ is a
plane-wave with wavevector $k_{zi}$. 
The full (i.e., energy and momentum dependent)
calculation of $V_{12}$ is a complicated task which, to our knowledge,
has not been undertaken so far. In this work we take the
$e_1$-$e_2$ interaction $V_{12}$ as the bare Coulomb
interaction, so that the solution of Eq.\ (\ref{diffeq}) is the
well-known Coulomb distortion factor\cite{calc}.
Note that the above approximations correspond to calculate the
two-electron function $M^{\inc}(\vR,z)$ in the free space; the
influence of the surface, however, is still present in the complete
wavefunction through the
single-electron functions $\psi^{\inc}(\vr_i)$ in Eq.\
(\ref{psi_fin})\cite{nota}. 

\narrowtext
We take $\Sigma_i(z_i)$ in a static approximation $V(z_i)$, with 
\beq
V(z) = \left\{\begin{array}{ll}
-U_0 / \left(A e^{Bz}+1\right) & z<0 \\ 
-(4z)^{-1} e^{-z/l_0} \left[1-e^{-\lambda z}\right] & z>0
\end{array}\right.
\label{jennings}
\eeq
which is the analytical form proposed in Ref.\ \onlinecite{Jones84}, 
with the inclusion of an exponential factor $e^{-z/l_0}$ 
characterized by a screening length $l_0$\cite{AB}. When $l_0\rightarrow
\infty$, $V(z)$ decays as the image potential at large $z$
(see inset in Fig.\ 1); inside the metal, $V(z)$ converges to the 
inner potential $U_0$.
As mentioned above, we assume that the initial state is the
unperturbed ground state of the He atom, which we describe by the
variational wavefunction\cite{LeSech94}
\beq
\langle\vr_1,\vr_2|\Psi_\ini\rangle = e^{-\sum_i r_i}
\left[1+\frac{r}{2}e^{-\alpha r}\right]
\sum_i \frac{\sinh\beta r_i}{\beta r_i},
\label{psi_ini}
\eeq
with variational parameters $\beta = 1.08 \mbox{a.u.}, \alpha = 0.15
\mbox{a.u.}$.

Figure 1 shows a polar representation of the calculated DCS with $ l_0
= \infty $. In order to better single out the effect of the surface, we
have chosen a configuration in which one electron is emitted at a large
polar angle with respect to the the normal to the surface; then, we
plot the DCS with the other electron emitted with a polar angle in the
range $(-\pi/2,\pi/2)$ and in a coplanar geometry; the condition $
\vk_1 + \vk_2 = 0$ is approximately met when both electrons are
emitted in the $z>0 $ half-space at large polar angles. We consider
linearly polarized light with the electric field parallel
(s-polarization) or perpendicular (p-polarization) to the surface. 

For comparison, we show in Fig.\ 1 the DCS calculated both with the
surface potential given by Eq.\ (\ref{jennings}) and with $ V(z)=0$
everywhere; note that the latter case, with no surface potential
present, is not fully equivalent to previous calculations\cite{DCS-He} of
the DCS for the free He, because the electron-ion interaction is
neglected in our calculation. Nonetheless, the shape of our calculated
DCS is in general agreement with those reported in Refs.\
\onlinecite{DCS-He}; in particular, the atomic selection rules
(\ref{sel_rule_1}) and (\ref{sel_rule_2}) are strictly obeyed.

\begin{figure} 
\noindent
\unitlength1mm
\begin{picture}(75,60) 
\put(0,5){\psfig{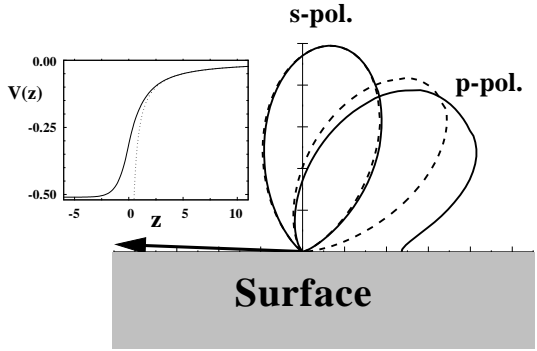}}
\end{picture}
\caption{Polar representation of the calculated DCS, normalized to the
maximum intensity. One electron is emitted with a large polar angle
$\theta\simeq 88^\circ$ with respect to the normal to the surface, as
indicated by the arrow. We consider s- and p-polarized light (electric
field parallel and perpendicular to the surface, respectively). Solid
lines: full calculation with the surface potential given by Eq.\
\ref{jennings} with $l_0=\infty$. Dotted lines: calculations with with
$V(z)=0$. Parameters are as follows: $E_1=E_2=10.5 $ eV,  
corresponding to a photon energy equal to 100 eV; 
$d=6.5 \mbox{a.u.}$. Inset:
potential profile, Eq.\ (\protect\ref{jennings}), used in the full
calculation (solid line), compared with the classical image potential
(dotted line). The parameters entering the potetial are $U_0=0.51
\mbox{a.u.}$, $\lambda = 1.25 \mbox{a.u.}$.}
\end{figure}

In the s-polarization the selections rules which apply in the atomic
case, which inhibit simultaneous emission of $e_1$ and $e_2$ in the
same as well as in opposite directions, are still strictly obeyed in
presence of the surface. In general, there is very little difference
between the full calculation and the $V(z)=0$ case. This is at
difference with the p-polarization; in this case, in fact, the surface
changes the DCS and shifts photo-emission towards larger polar angles;
note, in particular, that {\em 
the emission in opposite directions is
allowed}, in sharp contrast with the atomic case. 

The relaxation of the atomic selection rules in  the physisorbed system
is due to the long-range nature
of the surface potential. In Fig.\ 2 we show the calculated DCS for
selected values of $l_0$, from infinity to $l_0$ comparable to or smaller
than $d=6.5 \mbox{a.u.}$. The different curves are similar
to each other (and to the $V(z)=0$ case) in the s-polarization. For
p-polarization, instead, the photo-electron distribution is narrower
for small values of $l_0$. Moreover, the atomic selection rule $\vk_1 +
\vk_2 \neq 0$ is relaxed only as long as $l_0 \gg d$; the DCS at
large polar angles, in fact, drops rapidly when $l_0$ becomes
comparable to the physisorption distance; at the same time, for the
lowest values of $l_0$ the DCS develops modulations, and becomes
peaked along selected directions, corresponding to interference
between electrons emitted directly in the free space and those which
reach the vacuum after a reflection due to the surface potential.

\begin{figure} 
\noindent
\unitlength1mm
\begin{picture}(65,60) 
\put(10,5){\psfig{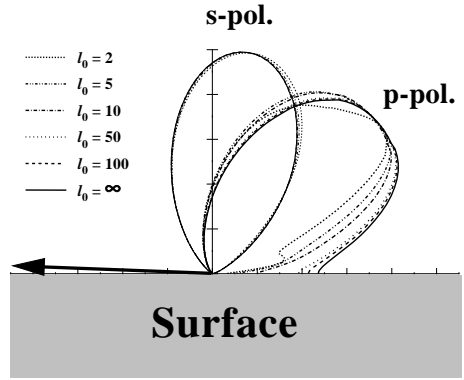}}
\end{picture}
\caption{Same as in Fig.\ 1 for selected values of $l_0$, given in
a.u. in the legend. All parameters are the same as in Fig.\ 1}
\end{figure}

In conclusion, we have analized the modifications induced on the DPI
of He by a polarizable surface. We have shown that the DCS depends on
the screening properties of the surface; in particular, the presence
of a long-ranged surface potential is associated with a substantial
relaxation of the atomic selections rules. On the other hand, the
precise angular distribution of the photoelectrons depends on the
details of the surface, such as the precise shape of the potential or
the physisorption distance. Therefore, DPI from metals promises to
become an interesting tool in assessing the screening properties of
the surfaces.

It is a pleasure to thank Carlo Calandra for several enlightning
discussions and a careful reading of the manuscript. Discussions with
Gianni Stefani are also gratefully acknowledged.

\end{multicols} 

\begin{references}

\bibitem{DCS-He} O. Schwarzkopf, B. Kr\"assig, J. Elminger, and V. Schmidt,
Phys. Rev. Lett. {\bf 70}, 3008 (1993); R. D\"orner {\em et al.}, 
Phys. Rev. A {\bf 57}, 1074 (1998).

\bibitem{Berakdar98} J. Berakdar, Phys. Rev. B {\bf 58}, 9808 (1998).


\bibitem{Osgood97} For a recent review, see R. M. Osgood and X. Wang, Solid
State Physics, {\bf 51}, 1 (1997).

\bibitem{gauge} For atomic He, previous calculations have shown that
for equal energy sharing the DCS calculated with the velocity and the
length gauge give very similar results (while the absolute
cross-section may differ). Moreover, the velocity gauge gives better
agreement with experiments when the two gauges differ quantitatively,
such as for unequal energy sharing \protect\cite{Maulbetsch94}.

\bibitem{UES} These selection rules do not hold if $E_1\neq E_2$
\protect\cite{Maulbetsch94}.

\bibitem{Eriksson85} H. G. Eriksson, B. R. Karlsson, and K. A. I. L.
Wijewardena, Phys. Rev. B {\bf 31}, 843 (1985).

\bibitem{deAbajo92} F. J. Garc\'\i a de Abajo and P. M. Echenique,
Phys. Rev. B {\bf 46}, 2663 (1992).

\bibitem{nota3C} We can make a comparison with the so-called 3C
wavefunctions used in three-body scattering
problems [C. R. Garibotti and J. Miraglia, Phys. Rev. A {\bf 21}, 572
(1980); M. Brauner, J. S. Briggs, and H. Klar, J. Phys. B {\bf 22},
2265 (1989); F. D. Colavecchia, G. Gasaneo, and C. R. Garibotti, Phys.
Rev. A {\bf 57}, 1018 (1998)] and, in particular, in the calculation of the DPI
cross-section of atoms (see Ref.\ \onlinecite{calc}); here, 
$\psi^\inc_{\alpha_i}(\vr_i)$ is determined by the single-particle surface
potential $V(z_i)$ rather than by the ion core Coulomb potential, as
in the above mentioned calculations.

\bibitem{calc} F. Maulbetsch and J. S. Briggs, Phys. Rev. Lett. {\bf
68}, 2004 (1992); L. R. Andersson and J. Burgd\"orfer, {\em ibid.}
{\bf 71}, 50 (1993); Y. Qiu, J.-Z. Tang, J. Burgd\"orfer, and J. Wang, 
Phys. Rev. A {\bf 57}, R1489 (1998).

\bibitem{nota} This is in analogy to what is usually done in the
calculation of the DPI cross-section of atoms, where the
two-electron distortion factor $M^\inc(\vr)$ is calculated neglecting
terms which arise from the bare electron-ion interaction\cite{calc}.
See also note \onlinecite{nota3C}.

\bibitem{Jones84} R. O. Jones, P. J. Jennings, and O. Jepsen, 
Phys. Rev. B {\bf 29}, 6473 (1984); P. J. Jennings, R. O. Jones, and
M. Weinert, {\em ibid.} {\bf 37}, 6113 (1988).

\bibitem{AB} The matching conditions for $V(z)$ and $dV(z)/dz$ at $z=0$
give the costants $A = 4 U_0/\lambda-1$, $B=(\lambda/2+1/l_0)
(4U_0/\lambda A)$. 

\bibitem{LeSech94} C. Le Sech, G. Hadinger, and M. Aubert-Fr\'econ, Z.
Phys. D {\bf 32}, 219 (1994).

\bibitem{Maulbetsch94} F. Maulbetsch and J. S. Briggs, 
J. Phys. B: At. Mol. Opt. Phys. {\bf 27}, 4095 (1994)

\end{references}
\end{document}